\documentclass[12pt]{article}

\usepackage{graphicx}

\begin{document}

\title{Fully polarized states and decoherence}

\author{Marco Frasca \\
        Via Erasmo Gattamelata, 3,\\
        00176 Roma (Italy)}

\date{\today}

\maketitle

\abstract{
The aim of this review is to show how ``ferromagnetic'' states, that is, states
having a fully polarization, can produce intrinsic decoherence by unitary evolution.
This effect can give an understanding of recent experiments on mesoscopic devices
as quantum point contacts showing the 0.7 conductance anomaly and the wide number
of data about saturation of dephasing time observed at very low
temperatures, as a fully polarized two dimensional electron gas. But similar
effects can be seen in different area of physics as for example the Dicke model
describing the interaction of two-level systems with a radiation mode. In this case one
can show that decoherence is intrinsic and remove a Schr\"odinger cat state leaving
a single coherent state, collapsing the wave function in the thermodynamic limit.
So, saturation of dephasing time at low temperatures in mesoscopic
devices can be understood by a fully polarized two dimensional electron gas that,
by an exchange model, can be reduced to a generalized form of the Dicke Hamiltonian and
where the quasiparticles are spin excitations interacting with magnons. In this way,
one can see that several experiments on nanowires and quantum dots can be
satisfactorily explained. The existence of intrinsic decoherence in the thermodynamic limit could have
deep implications in fundamental problems like quantum measurement and irreversibility.
Recent experiments with cavities with a large
number of photons and with nuclear magnetic resonance in organic molecular
crystals give a first strong support to this view.
}

\section{Introduction}

Decoherence, intended as the disappearance of interference terms and the emerging of the
classical world by the unavoidable interaction of a quantum systems with the degrees of
freedom of an external environment, is gaining increasing acceptance as a paradigm to
understand how classicality emerges from a quantum world \cite{zur1}. Notwithstanding
the increasing acceptance of this ideas, several criticisms have been put forward \cite{crit}.
Particularly, the common belief that decoherence could resolve the measurement problem
in quantum mechanics cannot be supported. Besides, the environment is quantum itself and all
the times we work out realistic examples, as we will see below, quantum electrodynamics with
a large number of degrees of freedom is implied. So, one may ask if the emergence of classicality
and the lost of quantum coherence is rather an intrinsic effect of unitary evolution in the
proper limits.

Recent experiments on mesoscopic devices \cite{mjw,nat,fer,mar} seem to indicate that
a new mechanism is at work producing decoherence at very small temperature. At present,
it is still an open question if the effect is an intrinsic one \cite{sac1,mw,sac2}. Besides,
an anomaly in the conductance in quantum point contacts, the so called 0.7 anomaly, has
prompted several people to put forward a hypothesis of a polarized electron gas \cite{wb,hw,hlw,bru}.
The same mechanism may be at work in the former case \cite{fra1,fra2}. A further decoherence
effect in quantum dots, due to hyperfine interaction between electrons and nuclei, has
been predicted \cite{loss1,loss2,loss3} assuming a fully polarized state for nuclei.

The existence of polarized states in mesoscopic devices has recently been supported by an
experiment on a two-dimensional electron gas (2DEG) \cite{ghosh}. Analysis based on numerical computations
were done about \cite{bach} showing the possible existence of such a state in 2DEG. Theoretical
studies also supported the existence of a ferromagnetic state 
in nanowires \cite{delin1,delin2,delin3}, that could be favored by the low density.

The relevance of polarized states to determine the very behavior at zero temperature of
mesoscopic devices could indeed reveal itself fundamental to understand a lot of issues
about the foundations of quantum mechanics. The question to be asked is if decoherence
should be considered as an effect of quantum evolution and if the ``collapse'' of the
wave function could be considered a new physical effect. A recent proposal for an experiment
to clarify this situation has been recently put forward \cite{pen}. The aim is to
see whether some stochastic effect is at work producing the ``collapse''. 

Recently, we pointed out how quantum electrodynamics is able to remove quantum coherence
in the thermodynamic limit \cite{fra3,fra4,fra5,fra6,fra7,fra8,fra9}. Two relevant assumptions are
at foundation of this conclusion. Firstly, we take a fully polarized state of two-level
atoms. Secondly, we take the thermodynamic limit. Both these assumptions are essential for
the proper working of an intrinsic mechanism of decoherence. If such a mechanism does
exist one can draw conclusions on the problem of quantum measurement and irreversibility.

Quantum measurement problem is a basic one in quantum mechanics and appeared since the
start with the Copenaghen interpretation. The essential point relies on the boundary between
a quantum and a classical word. That this boundary could be taken to infinity has been
shown recently with some experiments realized by Haroche's group \cite{har1,har2}. These
experiments confirm an understanding of quantum measurement in the thermodynamic limit,
taking the number of photons increasing, puts forward by Gea-Banacloche \cite{gb1,gb2} and
further analyzed by Knight and Phoenix \cite{kn1,kn2}. These experiments and theoretical
analysis seem to give a first hint toward the relevance of the thermodynamic limit in a
real understanding of the measurement problem in quantum mechanics.

The existence of an arrow of time is a basic problem in physics as also the Schr\"odinger equation
is reversible in time. This appears an old problem since the controversy between Boltzmann
and Loschmidt. Our conclusions rely on the assumption that a many-body effect in the
thermodynamic limit is acting removing reversibility. Essentially, our argument is based on the
very existence of singular limits in time, that is, periodic functions having fastly oscillating
phases in time are averaged away being blurred. Singular limits have been pointed out by Berry \cite{berry}
for the transition from a quantum to a classical world and pioneered, for the problem of
measurement in quantum mechanics, by Bohm \cite{bohm}.
 
A strong experimental evidence for this instability, proper
to many-body quantum systems, has been shown in NMR experiments with organic molecular crystals 
by Pastawski, Usaj and Levenstein\cite{pasta0,pasta1,pasta2,pasta3,pasta4}. These authors
were able to prove that the decoherence rate dependends just on the coupling constant between spins and
their number. The decoherence appears as an intrinsic effect as expected by our approach
with the proper dependencies and is rapidly attained in the thermodynamic limit, 
while the decay deviates from ordinary exponential to gaussian.

So, polarized states could be essential for the proper understanding of fundamental issues
in physics, relying the solution on quantum mechanical interactions between spins or
between two-level systems and bosonic modes.

The paper is structured in this way. In sec.\ref{sec2} we give an introduction to decoherence
by unitary evolution in the thermodynamic limit. 
In sec.\ref{sec3} we introduce the Dicke model and we analyze it both for a larger number
of bosons and fermions. 
In sec.\ref{sec4} we consider two exchange models for
a conduction electron into a polarized background and the Holstein-Primakoff transformation. 
In sec.\ref{sec5} we discuss the implications of dynamical decoherence
for the problem of irreversibility and the theory of quantum measurement. In sec.\ref{sec6}
the conclusions are given.

\section{Decoherence in the thermodynamic limit}
\label{sec2}

There is a deep relation between quantum mechanics and statistical mechanics. A first and
well-known connection is between the time evolution operator $e^{-iHt}$ ($\hbar=1$) and
the density matrix $e^{-\beta H}$, being $H$ the Hamiltonian of the system. The simple 
change $t\leftrightarrow -i\beta$ gives the connection. So, one may ask how far such a
relation could be taken. In statistical mechanics, a privileged role is played by the
thermodynamic limit to recover the laws of thermodynamics \cite{kada}. This limit is generally
taken by having the number of particles and volume going to infinity and keeping the density
constant. Anyhow, for some systems, the limit of the number of particles going to infinity is enough.

One may think to extend the concept of thermodynamic limit to quantum mechanics assuming that,
instead of the laws of thermodynamics, this limit recovers classical mechanics. Indeed, one
can prove the following theorem, for non-interacting quantum systems as happens for a perfect
gas in thermodynamic \cite{fra5}:

\newtheorem{classical}{Theorem}
\begin{classical}[Classicality]
An ensemble of N non interacting quantum systems, for properly chosen initial
states, has a set of operators $\{A_i,B_i,\ldots:i=1\ldots N\}$ from which one can derive a set
of observables behaving classically.
\end{classical}

A deep conceptual similarity is given through this theorem between statistical mechanics
and quantum mechanics for non-interacting systems. The proof is given by showing that, for a
proper chosen initial state, the quantum fluctuations are negligible with respect to the
average values of the observables that evolves in time in the thermodynamic limit. Two points
are essential to the proof: The thermodynamic limit and the proper chosen initial state.

Decoherence is obtained, after the theorem above, once the following statement is proved:

\begin{classical}[Decoherence]
An ensemble of N non interacting quantum systems,
having a set of operators $\{A_i,B_i,\ldots:i=1\ldots N\}$, from which one can derive a set
of observables behaving classically, interacting with a quantum system
through a Hamiltonian having a form like $V_0\otimes\sum_{i=1}^N A_i$, can produce
decoherence if properly initially prepared in a fully ``polarized'' state.
\end{classical}     

To prove this theorem we consider a Hamiltonian like
\begin{equation}
    H = H_S + \sum_{i=1}^N H_i + V_0\otimes\sum_{i=1}^N A_i 
\end{equation}
being $H_S$ the Hamiltonian of the quantum system, $H_i$ the Hamiltonian of the $i$-th
system in the bath and $V_0$ the interaction operator with the observables $A_i$ of
the bath. Then, passing to the interaction picture for the bath one has
\begin{equation}
    H_I = H_S + NV_0\otimes\sum_{i=1}^N\hat A(t)
\end{equation}
having introduced the operator
\begin{equation}
    \hat A(t) = {1\over N}\sum_{j=1}^N e^{iH_jt}A_je^{-iH_jt} 
\end{equation}
and we assume to exist the limit $N\rightarrow\infty$ for this operator at least on the chosen
initial state of the bath $|\chi\rangle$. This state is a fully polarized state as we take all
the systems in the bath in the same eigenstate of the Hamiltonian $H_i$. These two hypothesis
are essential for our proof.

Then, if the thermodynamic limit $N\rightarrow\infty$ is formally taken on the Hamiltonian $H_I$,
we enter a strongly coupling regime \cite{fra10,fra11,fra12,fra13} having at the leading order the
operator
\begin{equation}
     H_I' = NV_0\otimes\hat A(t)
\end{equation}
that can be diagonalized by the following unitary transformation
\begin{equation}
      U_0(t)=\sum_ne^{-iNv_n\bar at}|v_n\rangle\langle v_n||\chi\rangle\langle\chi|
\end{equation} 
being $V_0|v_n\rangle=v_n|v_n\rangle$ having assumed a discrete spectrum for the sake of
simplicity and $\bar a=\langle\chi|\hat A(t)|\chi\rangle=constant$. Then, the density matrix
of the quantum system in the bath becomes
\begin{equation}
      \rho_S(t)=\sum_n|\langle v_n|\psi_S(0)\rangle|^2 |v_n\rangle\langle v_n|+
	            \sum_{n\neq m}\langle v_m|\psi_S(0)\rangle\langle\psi_S(0)|v_n\rangle |v_n\rangle\langle v_m|
				e^{-iN(v_n-v_m)\bar at}.
\end{equation}
being $|\psi_S(0)\rangle$ the initial state of the quantum system.
To complete the proof, in the thermodynamic limit, a meaning should be attached to the
oscillating terms in the interference contribution in the density matrix. Here we face an
interesting problem. In order to recover an oscillating term we should be able to proper
sample it. Otherwise, if the sampling happens with a too small frequency, the numerical
rounding on the phases changes it into noise due to the large time step. 
This matter can be easily derived from the sampling
theorem \cite{opp}. This means that as $N$ increases to infinity, to recover the oscillating terms
becomes increasingly difficult as we are going to probe even more smaller time scales and the numerical
precision needed in time step to recover the phases is even more demanding. 
So, in the limit, we get just a noisy term 
with null average added to the mixed part of the density matrix and the interfering
part of the density matrix gives no contribution in the thermodynamic limit, completing
the proof of the theorem.

The conclusion to be drawn from the above about fastly oscillating functions is that
{\sl if a function has even more smaller time scales, e.g. due to the large number of particles,
phases are scrambled due to the need of a large sampling frequency, 
and a noisy behavior with a null average is obtained}.  

\begin{figure}[t,b,p]
\begin{center}
\includegraphics[height=0.5\textwidth,width=1\textwidth]{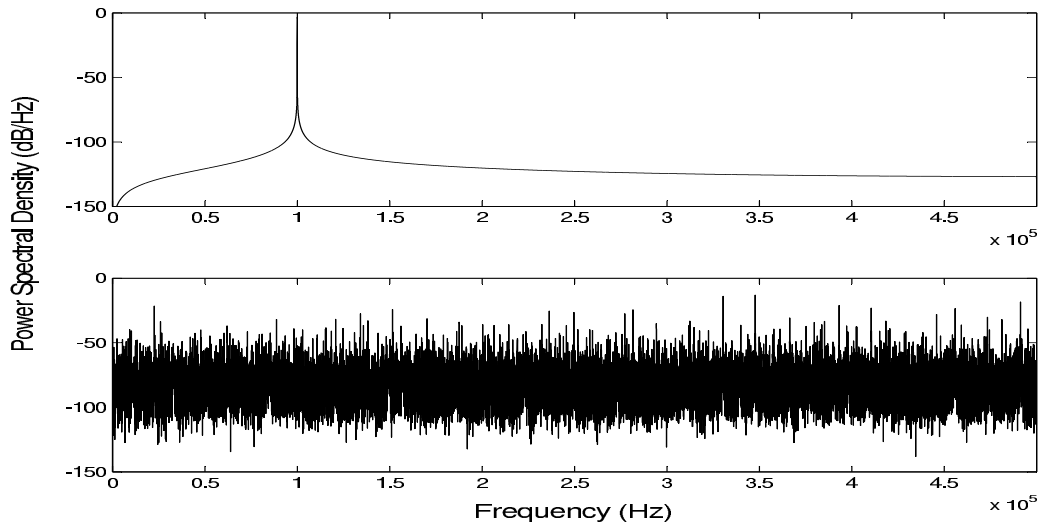}
\caption{\label{fig:fig1} Periodograms of the function $\sin(2\pi N t)$ with $N=10^5$ Hz (upper) 
and $N=10^{43}$ Hz (lower) both sampled with a frequency of 1 MHz.}
\end{center}
\end{figure}

Two further consideration are needed here. Firstly, we have provided another way to produce
a random sequence of numbers (see fig.\ref{fig:fig1}).
Secondly, one may ask, from a physical standpoint, if a meaning can be attached to unitary evolution
when this is probing Planck time scales. E.g., loop quantum gravity \cite{rov} does not support
unitary evolution at Planck time scale.

The impossibility to recover completely phases by a quantum system with a large number of
particles grants irreversibility as an intrinsic effect. The above theorems grant that also
decoherence is intrinsic. So, one appears the cause of the other.

\section{Decoherence in a bosonic-fermionic model}
\label{sec3}

One of the fundamental models in quantum electrodynamics, describing the interaction between
a single radiation mode with an ensemble of two-level atoms, is the Dicke model \cite{dicke}
described by the Hamiltonian
\begin{equation}
     H = \Delta S_z + \omega a^\dagger a + 2g S_x (a + a^\dagger)
\end{equation}
being $\Delta$ the separation between the two levels of the atoms, $a^\dagger$ and $a$ the
creation and annihilation operators for the radiation mode with frequency $\omega$, $g$ the
coupling constant that depends on the volume where the radiation stays as $V^{-{1\over 2}}$
and
\begin{eqnarray}
    S_x &=& \frac{1}{2}\sum_{n=1}^N\sigma_{xi} \\ \nonumber
    S_z &=& \frac{1}{2}\sum_{n=1}^N\sigma_{zi}
\end{eqnarray}
with $\sigma_{xi}$ and $\sigma_{zi}$ the Pauli matrices for the i-th atom.

We analyzed this model in the strong coupling regime \cite{fra3,fra6}. On this basis we showed
that the leading order should be given by
\begin{equation}
    H_0 = \omega a^\dagger a + 2g S_x (a + a^\dagger)
\end{equation}
that can be immediately diagonalized giving the time evolution operator
\begin{equation}
\label{eq:u0}
     U_0(t) = e^{i\frac{4S_x^2g^2}{\omega^2}(\omega t-\sin(\omega t))} 
	 e^{2S_x[\beta(t)a^\dagger-\beta^*(t)a]}
\end{equation}
being $\beta(t) = \frac{g}{\omega}(1-e^{i\omega t})$ and treating $S_x$ as a c-number. It is
immediately realized that, for a fully polarized state $|\chi\rangle$ such that 
$S_x|\chi\rangle=-\frac{N}{2}|\chi\rangle$ and the radiation mode in the ground state $|0\rangle$
one has, in the thermodynamic limit $N\rightarrow\infty$ and constant volume, a coherent state for
a classical radiation field
\begin{equation}
     |\psi\rangle = e^{i\frac{N^2g^2}{\omega^2}(\omega t-\sin(\omega t))}|-N\beta(t)\rangle|\chi\rangle.
\end{equation}
A polarized state for the atomic ensemble has produced the effect to amplify the quantum fluctuations
of the ground state of the radiation mode to the classical level, an effect we called 
quantum amplifier (QAMP). The state of the ensemble of two-level atoms is left untouched and
acts passively. Higher order corrections where also given \cite{fra4,fra6,fra9}. 

Results about the classicality in the thermodynamic limit of
observables as the number operator and ``spin'' 
in the Dicke model were given in pioneering papers by Hepp and Lieb \cite{hl1,hl2}. So,  
classicality seems a rather general property of this model than the strong coupling approximation seems to grant. 
Recently, an in-depth study by Brandes and Emary \cite{be1,be2}, proved that the Dicke model undergoes
a quantum phase transition. These authors considered the limit with the number of particles and
the volume going to infinity and keeping the density constant. In this case, one can introduce
the coupling parameter as $g = \frac{\lambda}{\sqrt{N}}$ and the thermodynamic limit just amounts
to take the limit $N\rightarrow\infty$, remembering that $g$ depends on the volume as $V^{-{1\over 2}}$
with the Hamiltonian being
\begin{equation}
     H = \Delta S_z + \omega a^\dagger a + 2\frac{\lambda}{\sqrt{N}} S_x (a + a^\dagger).
\end{equation}
The quantum phase transition happens in the parameter space at $\lambda_c=\frac{\sqrt{\Delta\omega}}{2}$.
To reach their aim, Brandes and Emary used the Holstein-Primakoff transformation \cite{hp} given by
\begin{eqnarray}
    S_+ &=& \sqrt{N}b^\dagger\left(1-\frac{b^\dagger b}{N}\right)^{\frac{1}{2}} \\ \nonumber
	S_- &=& \sqrt{N}\left(1-\frac{b^\dagger b}{N}\right)^{\frac{1}{2}}b \\ \nonumber
	S_z &=& b^\dagger b - \frac{N}{2}
\end{eqnarray}
that, assuming a maximal spin state, gives a series in $\frac{1}{N}$ with the Hamiltonians
at various orders
\begin{eqnarray}
    H_0 &=& \Delta b^\dagger b + \omega a^\dagger a + \lambda(a^\dagger + a)(b^\dagger +b) \\ \nonumber
	H_1 &=& -\frac{\lambda}{2N}(b^\dagger b^\dagger b + b^\dagger bb) \\ \nonumber
	&\ldots&.
\end{eqnarray}
It easy to see that the leading order Hamiltonian, describing the behavior of the
model in the thermodynamic limit, can be easily diagonalized. So, in this limit, the
Dicke model is exactly integrable. These authors reached two relevant conclusions that are
fundamental for our aims. Firstly, they showed that any macroscopic coherence disappears in 
the thermodynamic limit. Secondly, by a numerical effort, they proved that,
keeping $N$ fixed and letting the coupling
constant increase as $\lambda\rightarrow\infty$, the model is well described by the Hamiltonian
\begin{equation}
    H_{lim} = \omega a^\dagger a + 2\frac{\lambda}{\sqrt{N}}S_x(a^\dagger + a)
\end{equation}
with the proper quantization axis being given by $S_x$. This latter result gives a strong
support to our above work. But this conclusions can be pushed further and to prove 
analitically that \cite{fra9}, also at constant volume, {\sl in the limit of $N\rightarrow\infty$ or 
$g\rightarrow\infty$ or both, the Hamiltonian $H_{lim}$ becomes exact and the Dicke model
becomes integrable again.} This can be proved straightforwardly by applying the Holstein-Primakoff
transformation, this time, with respect to $S_x$ axis as
\begin{eqnarray}
    S_x &=& -\frac{N}{2}+c^\dagger c \\ \nonumber
	S_+ &=& \sqrt{N}c^\dagger\left(1-\frac{c^\dagger c}{N}\right)^\frac{1}{2} \\ \nonumber
	S_- &=& \sqrt{N}\left(1-\frac{c^\dagger c}{N}\right)^\frac{1}{2}c.
\end{eqnarray}
being $S_z = \frac{1}{2}(S_++S_-)$.
We note that this transformation cannot be applied in the problem studied by Brandes and Emary
while their approach cannot be applied here. These methods just complete each other with the
boundary being given by the Hamiltonian $H_{lim}$. One has in this way, turning back to
the coupling constant $g$,
\begin{equation}
    H = \omega a^\dagger a-Ng(a+a^\dagger) + 2gc^\dagger c (a+a^\dagger)+
	\frac{\Delta}{2}\sqrt{N}
	(c^\dagger+c)-\frac{\Delta}{4}\frac{1}{\sqrt{N}}(c^\dagger c^\dagger c + c^\dagger cc)+\ldots.
\end{equation}
that is not easy to manage unless the unitary transformation
\begin{equation}
    U_0 = e^{\frac{2g}{\omega}c^\dagger c(a-a^\dagger)}
\end{equation}
is applied giving the Hamiltonian at the various orders
\begin{eqnarray}
    H'_0 &=& \omega a^\dagger a - Ng(a+a^\dagger) + 4N\frac{g^2}{\omega}c^\dagger c \\ \nonumber
	H'_1 &=& \frac{\Delta}{2}\sqrt{N}\left[c^\dagger e^{-\frac{2g}{\omega}(a-a^\dagger)}
	+c e^{\frac{2g}{\omega}(a-a^\dagger)}\right] \\ \nonumber
	H'_2 &=& - 4\frac{g^2}{\omega}(c^\dagger c)^2 \\ \nonumber
	H'_3 &=& - \frac{\Delta}{4N}\left[c^\dagger c^\dagger c e^{-\frac{2g}{\omega}(a-a^\dagger)}
	+c^\dagger cc e^{\frac{2g}{\omega}(a-a^\dagger)}\right] \\ \nonumber
	&\vdots& 
\end{eqnarray}
proving our initial assertion for $N\rightarrow\infty$. The prove for $g\rightarrow\infty$ 
can be obtained directly by the Rayleigh-Schr\"odinger series \cite{fra9}.

The main question to ask now is if the Hamiltonian $H_{lim}$ can indeed provide decoherence.
The point here is delicate and involves again the concept of a singular limit. As a matter
of fact, we already showed that can generate classical state of radiation field producing
a new effect we named QAMP. But let us consider a phase Schr\"ondiger cat state 
for the radiation field defined as\cite{sch}
\begin{equation}
    |\zeta\rangle = {\cal N} (|\gamma e^{i\phi}\rangle + |\gamma e^{-i\phi}\rangle)
\end{equation}
being $\gamma$ and $\phi$ two real parameters, $\cal N$ a normalization constant. These are
two superposed coherent states of the radiation field. So, we can apply the unitary evolution
(\ref{eq:u0}) to this state obtaining
\begin{equation}
     |\psi(t)\rangle = e^{i\xi(t)}{\cal N}(e^{i\phi_1(t)}
	 |\frac{Ng}{\omega}(e^{i\omega t}-1) 
	 + \gamma e^{i\phi-i\omega t}\rangle 
	 + e^{i\phi_2(t)}|\frac{Ng}{\omega}(e^{i\omega t}-1) + \gamma e^{-i\phi-i\omega t}\rangle)|0\rangle_-
\end{equation}
where a ground state has been assumed for the other bosonic mode built from the two-level atoms.
This result appears rather surprising as, taking formally the limit $N\rightarrow\infty$,
the superposition is removed leaving just a classical state of radiation. This could claimed only
if the quantum effect due to the initial state would be really removed and not just displaced
to infinity or hidden by the level of description. Indeed, infinity is not a really comfortable
place to do physics and this can be seen by considering the interfering part of the Wigner
function of the above state, given by
\begin{eqnarray}
    W_{INT} &=&\frac{2}{\pi}
	\exp\left[-\left(x + \frac{\sqrt{2}Ng}{\omega}(1-\cos(\omega t)) - 
	\sqrt{2}\gamma\cos(\phi)\cos(\omega t)\right)^2\right] \\ \nonumber
	&\times&\exp\left[-\left(p + \frac{\sqrt{2}Ng}{\omega}\sin(\omega t) + 
	\sqrt{2}\gamma\cos(\phi)\sin(\omega t)\right)^2\right] \\ \nonumber
	&\times&\cos\left[2\sqrt{2}\gamma\sin(\phi)\left(p\sin(\omega t) - x\cos(\omega t)\right)
	+\gamma^2\sin(2\phi) + 
	8\gamma\frac{Ng}{\omega}\sin(\phi)(1-\cos(\omega t))\right].
\end{eqnarray}
and we easily recognize a singular limit as happened into the prove of the theorem on
decoherence in the thermodynamic limit. Again we can get noise with null average 
in the thermodynamic limit and we have true decoherence.

The relevance of the above analysis relies on the fact that this study implies a realistic
model of the radiation-matter interaction that has wide applicability. This takes us to the
important conclusions that electromagnetic interaction is the main vehicle of decoherence and,
being also the means used to do a measurement, we get an understanding of why quantum
coherence gets lost in a measurement device. It is essential to point out again that our
conclusions rely on the two fundamental hypothesis of a fully polarized state and the
thermodynamic limit.    

\section{Decoherence in an exchange model}
\label{sec4}

The discovery of weak localization \cite{alt} permitted to realize some experiments to measure the
coherence time in mesoscopic devices. First experiments pointed out that, differently from what
the theory predicts, a saturation of the dephasing time was seen \cite{sat1,sat2,sat3}. Initially,
this was not considered a concern as some external effects could easily explain the saturation.
In 1997 an experiment by Mohanty, Jariwala and Webb, realized in order to eliminate any possible 
external effects, proved that the saturation of the dephasing time seems to be intrinsic. This
started a debate that is still open \cite{sac1,mw,sac2} with the fundamental question to
be answered if the effect is really intrinsic or not. This is a very fundamental question as
hits the possibility to realize fully coherent mesoscopic devices. 

Curiously enough, the debate considers just some devices neglecting e.g. the situation on quantum dots
where one has two contrasting experiments \cite{fer,mar} but both confirming the saturation. The
discrepancy between the two experiments relies on the fact that in \cite{fer} was observed
a dependence of saturation time on the number of electrons in the two dimensional electron gas
of the dot, while in \cite{mar} this dependence was not seen. But it should be said that different
approaches were used to interpret the data. If a dependence on the number
of electrons exists then this also means a dependence on the size, being the density kept constant.

A dependence on the size for the saturation of dephasing time in nanowires has been also
experimentally observed \cite{nat}. While this geometrical dependence should be confirmed
by further experiments, no experimental proposal seemed to address this point so far.
But some experimental  and numerical results in mesoscopic physics for 2DEG, 
nanowires, nanoclusters and so on \cite{ghosh,bach,delin1,delin2,delin3,nano} 
point out toward a possibility that the Fermi electron gas in this devices may be
polarized. Besides, a similar hypothesis was suggested for the 0.7 conductance anomaly \cite{wb,hw,hlw,bru}
even if an objection has been put forward here based on a theorem by Lieb and Mattis\cite{lm}.
A polarized Fermi gas can easily explain the dependence on geometry of the saturation time.

Let us see this in a simple model. We consider as a possible Hamiltonian \cite{fra1}
\begin{equation}
     H = \frac{\Delta}{2}\sigma_z - J\sigma_x\sum_{i=1}^N\tau_{xi}
\end{equation}
where we consider negligible the dynamical part of the $N$ spins that interact with a single one through
the coupling constant $J$. $\Delta$ is the separation between the energy levels of the single spin.
Both $\sigma$ and $\tau$ represent Pauli matrices. The sign of the coupling constant $J$ does not
matter in this case. This appears as a simplified exchange model describing the interaction
between a conduction electron and the Fermi gas. 
It is straightforward to write down the unitary evolution as
\begin{equation}
    U_0(t)=\exp\left[-it\left(
	       \frac{\Delta}{2}\sigma_z-J\sigma_x\sum_{i=1}^N \tau_{xi}
		   \right)\right]
\end{equation}
that can be disentangled as \cite{are}
\begin{equation}
    U_0(t)=\exp\left(-i\hat\Lambda(t)\sigma_+\right)
	       \exp\left(-\ln\hat\Sigma(t)\sigma_3\right)
		   \exp\left(-i\hat\Lambda(t)\sigma_-\right)
\end{equation}
being the operators
\begin{eqnarray}
    \hat\Lambda(t)&=&\frac{-J\sum_{i=1}^N\tau_{xi}}{\hat\Omega}
	                 \frac{\sin(\hat\Omega t)}{\hat\Sigma(t)} \\
    \hat\Sigma(t)&=&\cos(\hat\Omega t)+i\frac{\Delta}{2\hat\Omega}\sin(\hat\Omega t) \\
	\hat\Omega(t)&=&\sqrt{\frac{\Delta^2}{4}+J^2\left(\sum_{i=1}^N\tau_{xi}\right)^2}.
\end{eqnarray}
Assuming a fully polarized state, in the lower spin state, for the Fermi gas $|\chi\rangle$
and $N$ being enough large, one has the following wave function for the conduction electron
\begin{equation}
    |\psi(t)\rangle = \approx\exp\left(-itNJ\sigma_x\right)|\psi(0)\rangle.
\end{equation}
So, it is easy to see that, when the conduction electron is in the lower eigenstates of
$\sigma_z$ one gets a spin coherent state \cite{are}. But we are interested in the density
matrix of the conducting electron. This can be written as
\begin{eqnarray}
    \rho_{1,1}(t)&=&\frac{1-\cos(2NJt)}{2} \\
	\rho_{1,-1}(t) &=& -i\frac{1}{2}\sin{2NJt} \\
	\rho_{-1,1}(t) &=& i\frac{1}{2}\sin{2NJt} \\
	\rho_{-1,-1}(t)&=&\frac{1+\cos(2NJt)}{2}.
\end{eqnarray}
We can easily recognize that we can have oscillation with increasing frequency in the limit
$N\rightarrow\infty$. We have again a singular limit and the time scale for it is given by
\begin{equation}
    T_s = \frac{\pi}{NJ}.
\end{equation}
It should be noted at this point that, if the diffusion constant can be taken as $D\sim J$,
we recover the experimental results by Lin and Kao \cite{lin} but here the dependence on the
geometry was not tested. Finally, we have showed as a  simple exchange model with the
condition of a singular limit recovers the dependence on the number of electrons $N$ or,
keeping the density constant, on the geometry. The averaged density matrix permits us to
recover classical probabilities reducing the density matrix to a mixed form, i.e. we have decoherence.

We can make the above model more realistic and consider an exchange model like \cite{defm}
\begin{equation}
    H = H_0 + H_h + H_e
\end{equation}
being
\begin{equation}
    H_0 = \sum_{{\bf p}\sigma}E_{\bf p}c^\dagger_{{\bf p}\sigma}c_{{\bf p}\sigma}
\end{equation}
the Hamiltonian describing the itinerant electrons, 
\begin{equation}
    H_h = -J_h \sum_{\langle ij \rangle}{\bf S}_i\cdot{\bf S}_j
\end{equation}
is the Heisenberg term of ferromagnetic type, $J_h>0$, representing the interaction between the
spins of the gas, and
\begin{equation}
    H_e = J \sum_i{\bf S}_i\cdot{\bf s}_i
\end{equation}
is the exchange term (a Kondo term as from the first Hund's rule), being
\begin{equation}
    {\bf s}_i = \sum_{\alpha\beta}c^\dagger_{i\alpha}{\bf s}_{\alpha\beta}c_{i\beta}
\end{equation}
with ${\bf s}_{\alpha\beta}$ spin matrices whose components for spin $\frac{1}{2}$ are
given by $\frac{\mbox{\boldmath $\sigma$}_{\alpha\beta}}{2}$ with 
$\mbox{\boldmath $\sigma$}_{\alpha\beta}$ the Pauli matrices. The sign of the
coupling constant $J$ in the exchange term should be determined on a physical ground.
Besides, in order to favor the tendency of the conduction electron to align, 
we neglect $H_0$.

In order to study this model in a situation of a fully polarized electron gas, we use
again the Holstein-Primakoff transformation as
\begin{eqnarray}
    S^+_i&=&a^\dagger_i\left(2S-a^\dagger_ia_i\right)^\frac{1}{2} \\ \nonumber
	S^-_i&=&\left(2S-a^\dagger_ia_i\right)^\frac{1}{2}a_i \\ \nonumber
	S^+_i&=&S-a^\dagger_ia_i.
\end{eqnarray}
We arrive at the following expression, omitting $H_0$ as assumed initially,
\begin{equation}
    H' = -2zNJ_hS^2+\sum_{\bf k}\epsilon_{\bf k}a^\dagger_{\bf k}a_{\bf k}+
	JS\sum_i s^z_i + J\sqrt{\frac{S}{2}}\sum_{\bf k}
	\left(a^\dagger_{\bf k}s^-_{\bf k}+a_{\bf k}s^+_{\bf k}\right)
\end{equation}
being $\epsilon_{\bf k}=2zJ_hS(1-\gamma_{\bf k})$ and 
$\gamma_{\bf k}=\frac{1}{z}\sum_{\bf a}e^{i{\bf k}\cdot{\bf a}}$ with $\bf a$  the
vector linking two nearest neighbor spins and $z$ the number of nearest neighbor spins.
We recognize here the Dicke model seen in the preceding section, with the main difference
that we have a large number of modes and the spin operator depending on $\bf k$. This means
that the known considerations about radiation-matter interaction apply and we can have
Rabi oscillations for a small number of modes becoming a decaying exponential, i.e. an
effect of decoherence, when the number of modes becomes larger. Indeed, putting
\begin{equation}
    s^+_{\bf k}=\frac{1}{\sqrt{N}}\sum_i s^+_i e^{i{\bf k}\cdot{\bf r}_i}
\end{equation}
and similarly for $s^-_{\bf k}$, instead of itinerant electrons, we
have quasi-particles being spin excitations, described by the Hamiltonian
\begin{equation}
    H_S = JS\sum_i s^z_i = 
	\frac{JS}{2}\sum_{\bf k}\left(c^\dagger_{{\bf k}\uparrow}c_{{\bf k}\uparrow}
	-c^\dagger_{{\bf k}\downarrow}c_{{\bf k}\downarrow}\right), 
\end{equation}
interacting with magnons. Finally, passing to interaction picture we get
\begin{equation}
\label{eq:HI}
    H_I = J\sqrt{\frac{S}{2}}\sum_{\bf k}
	\left(a^\dagger_{\bf k}s^-_{\bf k}e^{i(\epsilon_{\bf k}-JS)t}+
	a_{\bf k}s^+_{\bf k}e^{-i(\epsilon_{\bf k}-JS)t}\right)
\end{equation}  
and we can immediately identify to the leading order the processes that
can induce decoherence, that is, we can have an itinerant electron to flip
its spin by emitting a magnon or, being a magnon present, by absorption. We can
conclude that the only possible choice for the coupling is $J>0$ proper to a Kondo
Hamiltonian. So, we can immediately apply the Fermi golden rule to the above Hamiltonian
obtaining the rate of spin flip
\begin{equation}
     \Gamma = 2\pi\frac{J^2S}{2}\sum_{\bf k}\delta(\epsilon_{\bf k}-JS).
\end{equation}
Passing from the sum to the integral, a crucial step, we get, for a two dimensional device,
\begin{equation}
    \Gamma_{d=2} = \frac{1}{2}Vm^*J^2S
\end{equation}
where the dependence on the geometry is evident in the volume $V$ and we have assumed
\begin{equation}
    \epsilon_{\bf k}=\frac{{\bf k}^2}{2m^*}
\end{equation}
for the dispersion relation, being $m^*$ the effective mass of the electron. We note that a 
proper experimental verification of this theory relies on the assumption that the diffusion
coefficient $D$ is kept constant during measurement as the quantity measured is the coherence
length $L_\phi$ related to the dephasing time $\tau_\phi$ by the relation $L_\phi=\sqrt{D\tau_\phi}$.
At the present, the only experiment that does this is the one by Natelson's group \cite{nat}
that gives the right behavior on geometry. Experiments on quantum dots give directly the
dephasing time but, as said above, contrasting results were obtained. 
Then, further experimental work is needed.
 
\section{Quantum measurement and irreversibility}
\label{sec5}

The Dicke model represents a truly representative description of radiation-matter interaction
in the non-relativistic limit. This model permits to prove that decoherence is intrinsic to
unitary evolution when two more conditions are taken: a proper initial state and the
thermodynamic limit. Besides, we have seen that recovering phases can be very demanding when
the thermodynamic limit is taken as the time scale becomes even more smaller. Sampling
such rapidly oscillating functions is not physically realizable and we have to consider
averages in time. This has the main effect to remove indeterminacy on Schr\"odinger cat
states as we have proved. Besides, superposition states are reduced to a single pure state.
This points to consider the radiation-matter interaction as the main vehicle for decoherence.
Indeed, it is really difficult to see a measurement device that does not use electromagnetic
interactions for its aims.

This implies that there exists an intrinsic instability in the determination of the phases in
quantum mechanics that is properly acting when the thermodynamic limit is taken. This effect
seems to have been seen in recent NMR experiment made by 
Pastawski, Levenstein and Usaj\cite{pasta0,pasta1,pasta2,pasta3,pasta4}. The aim of this
authors was rather different with respect to ours. Indeed, the existence of the irreversibility,
a rather common experience in the macroscopic world, is a longstanding problem in physics
that can be traced back to the controversy between Loschmidt and Boltzmann (see \cite{pasta0} and refs. therein). 
The essential point is that all the fundamental laws of physics appear to be reversible in time, so
whatever process one can realize in one direction of time, is perfectly legal when time is
reversed. Then, it is very difficult to see how, from such a symmetry, the existence of entropy
or of the Boltzmann $H$ function can be obtained. One could think to reverse all the velocities
of the particles and turn back to the initial state, reversing the natural order-disorder
sequence that is never observed. 

In order to solve this problem, Boltzmann introduced the statistical hypothesis of
``molecular chaos'' ({\sl Stosszahlansatz}) \cite{huang}, assuming that the reversing claimed by Loschmidt 
is not physically realizable. We know today that chaos inherent to Hamiltonian systems as
expressed by the Kolmogorov-Arnold-Moser theorem \cite{mos}, can give a proper answer at the level of
classical mechanics to the Boltzmann's hypothesis. But we also know that world is not classical
but quantum and we have to find a similar mechanism in quantum mechanics to support Boltzmann's
hypothesis and explain irreversibility. The problem we meet here is that chaos does not
exist in quantum mechanics, being the Schr\"odiger equation linear.

Experiments by Pastawski, Levenstein and Usaj have shown that, in NMR with organic
molecular crystals as ferrocene, cobaltocene and others, reversibility for this spin systems is not recovered
and the proper dephasing time, being the system unable to recover the initial phase, decreases
as the thermodynamic limit is taken, proving that the effect is even more important in this
limit. This is exactly the same behavior we have derived from both the exchange models in the preceding section. 
Another interesting result in their experiments is that the decay is gaussian, differently
from the exponential decay expected by ordinary decoherence effects.

An interesting point here is how physically realized is the thermodynamic limit. The explanation
is given by the existence of phase transitions in condensed matter. Statistical mechanics requires,
from a mathematical standpoint, the thermodynamic limit for the existence of phase transitions \cite{ons,yl}.
For all practical purposes, as our everyday experience proves, Avogadro number is enough to
have a thermodynamic limit.

Dephasing, as also shown in the Dicke model where it provokes decoherence in the thermodynamic
limit, is then at the root of the quantum measurement problem and irreversibility as initially
conceived by Boltzmann. Scrambling of rapidly oscillating phases due to the rough sampling 
is the cause of both the effects, being effectively noise.

\section{Conclusions}
\label{sec6}

The conclusions to be drawn are that both on experimental and theoretical sides there are
strong hints for the existence of a many-body effect that would permits to understand the
origin of irreversibility and implies decoherence in the unitary evolution.

The main assumptions we have made to recover such an effect is the thermodynamic limit and
a fully polarized state, that is, we need a proper initial preparation of the quantum system.

So, further confirmations of the relevance of the thermodynamic limit in quantum mechanics
would extend significantly the theory, removing a somewhat undefined concept of environment,
as conceived today. So, quantum mechanics should be seen as the
bootstrap program of our classical world.

\end{document}